\documentstyle[aps,epsf,epsfig]{revtex}
\draft

\newcommand{\cl}{\cline{2-7}}
\newcommand{\be}{\begin{equation}}
\newcommand{\ee}{\end{equation}}
\newcommand{\ba}{\begin{eqnarray}}
\newcommand{\ea}{\end{eqnarray}}
\newcommand{\bu}{\bullet}

\title{Masses and internal structure of mesons in the string quark model}
\author{L.D. Soloviev}
\address{ Department of Physics, University of Michigan, Ann Arbor, 
MI 48109-1120, USA,\\
  and\\
Institute for High Energy Physics, 142284, Protvino, Russia }
\date{20 February 1999}

\begin{document}
\maketitle

\begin{abstract}
The relativistic quantum string quark model, 
proposed earlier, is
applied to all mesons, from pion to $\Upsilon$, lying on the leading
Regge trajectories (i.e., to the lowest radial excitations in terms
of
the potential quark models). The model describes the meson mass
spectrum, and comparison with measured meson masses allows one to
determine the parameters of the model: current quark masses,
universal string tension, and phenomenological constants describing
nonstring short-range interaction. The meson Regge trajectories are
in general nonlinear; practically linear are  only trajectories for
light-quark mesons with non-zero lowest spins. The model predicts
masses of many new higher-spin mesons. A
new $K^*(1^-)$ meson is predicted with mass 1910~Mev. In some cases
the 
masses of new
low-spin mesons are predicted by extrapolation of the
phenomenological short-range parameters in the quark masses. In this
way the model predicts the mass of $\eta_b(1S)(0^{-+})$ to be 
$9500\pm 30$~MeV, and the mass of $B_c(0^-)$ to be $6400\pm 30$~MeV
(the potential model predictions are 100 Mev lower).  The
relativistic wave functions of the composite mesons allow one to
calculate the energy and spin structure of mesons. The average
quark-spin
projections in polarized  $\rho$-meson are twice as small
as the nonrelativistic quark model predictions. The spin structure of
$K^*$ reveals an 80\% violation of the flavour $SU(3)$. These
results
may be relevant to understanding the ``spin crises'' for nucleons.
\end{abstract}


\section{Introduction}
The naive quark model of hadrons, while attractive in its simplicity, is not
quite so simple under closer examination. It is not relativistic since it
contains a confinement potential proportional to a space distance.
One can introduce a quasipotential dependent on distance and
momentum,
which makes the wave equation Lorentz covariant, but then the
phenomenological quasipotential is not simple. The model contains
constituent quarks which are purely phenomenological notions. Their
masses are not fundamental and can vary from mesons to hadrons and
even from one meson to another. Their spins are not fundamental either; 
and the ``spin crises'' for nucleons suggest that the quark spins should be
different from
1/2, or that the naive quark model is too naive. 

In Refs. [1,2] an alternative, string quark model (SQM) has been
proposed
which  contains  neither potential nor constituent
quarks. The physical origin of  confinement and  constituent quarks, 
the gluon field, is taken into account
explicitly, in an approximation of the quantum Nambu-Goto string. The
string
provides a confinement mechanism and, since the string is a
physical object with its own energy-momentum and angular momentum,
the quarks at the ends of the string are fundamental quarks with
current masses and spin 1/2.

The application of SQM in Ref. [1] was confined to a particular type
of
leading meson Regge trajectories. Here we consider all four types of
them, obtain relativistic wave functions of composite mesons, and
calculate the internal (energy and spin ) structure of mesons.

As in Ref. [1], we consider  only the simplest string configuration
--- the
rotating straight line, which is responsible for the leading Regge
trajectories of mesons. The daughter trajectories (i.e., the higher
radial excitations in terms of potential models) correspond to
vibrations of the string.

The model is quantized in accord  with Poincar\'e invariance and,
due to account of quark spins, contains no tachyons.

The model predicts that the Regge trajectories for light-quark mesons
with lowest spin 1 ($\rho$-type and $b_1$-type) are practically
linear.
The corresponding trajectories for heavy-light-quark mesons are not
linear, but, to a good approximation, can be represented by straight
lines
for spins less than 6 by replacing the argument $m^2$ by $(m -
m_h)^2$,
where $m_h$ is the heavy-quark mass. The slopes of these straight
lines
are bigger than for the light-quark mesons, and increase with $m_h$,
the
limit value being twice as big as for the light-quark mesons. The
trajectories for heavy quarkonia are essentially nonlinear.

The Regge trajectories with lowest spin 0 ($\pi$-type and $a_0$-type)
are always nonlinear in the low-spin region.

The model describes masses of all mesons, from pion to $\Upsilon$,
lying on the leading Regge trajectories. The main parameters of the
model, the universal string tension and the current quark masses,
have
been determined in [1] by comparison with experimental meson masses
lying on the $\rho$-type trajectories. So, for each other trajectory
(without mixing), we have only one unknown short-range parameter.
Experiment suggests that these parameters for the $\pi$-type and the
$b_1$-type trajectories are equal. The short-range parameters do not
strongly depend on the quark masses, and in some cases can be
obtained
from known parameters by a safe extrapolation in the quark masses.

As a result, the model predicts masses and other quantum numbers of
many higher-spin mesons and some low-spin mesons. For instant, the
model predicts a
new $K^*(1^-)$ meson with mass 1910~Mev (without extrapolation) and
 the masses of $\eta_b(1S)(0^{-+})$ and $B_c(0^-)$ to be 
$9500\pm 30$~MeV (bigger than the $\Upsilon$-mass) and  $6400\pm
30$~MeV,
respectively. The corresponding predictions of a
 potential quark model (PQM) [5] are 100 Mev lower. This number can
characterize difference between many SQM and PQM predictions, so that
further systematic experimental study of meson spectrum with accuracy
capable to distinguish these predictions seems to be important for
understanding confinement.

The SQM relativistic wave functions of composite mesons allow one to
calculate the meson internal structure. The separate string and quark
contributions into meson masses are obtained. The average spin
projections
of  $u$- and $\bar d$-quark in polarized $\rho^+$, divided by the
same 
projection of the total meson spin, are found to be 0.22 and 0.23,
respectively, i.e., twice as small as the nonrelativistic quark model
prediction 0.5.

The corresponding numbers for  $u$- and $\bar s$-quark within
$K^{*+}$
are 0.22 and 0.42, respectively. This means that the flavour $SU_3$ 
is violated up to 80\%  for the spin structure in the relativistic
model.

The results on the meson spin structure suggest a new approach to
understanding the nucleon spin structure, obtained from polarized
deep inelastic lepton-nucleon scattering and extrapolated to low
$Q^2$
(the so called spin crises).

At the same time, the above numbers for the $\rho$ spin structure are
different from 0.17, the number corresponding to vanishing quark
masses, so 
that
one can hope that future polarization experiments will allow one to
estimate
the light-quark current masses from experiment.

The outline of the paper is the following. In Sec. 2, the string
physics
is described in the classical approximation. It follows, of course,
from
the string equations of Sec. 3, but can be described in familiar
terms
of the pointlike-particle mechanics, if only few basic properties of
the
string are taken from the equations. This description shows that the
string, presumably realized by the gluon field inside mesons, is
quite
a new object from mechanical viewpoint. Sec. 2 also clarifies the
origin
and properties of the string functions, relevant to the quantum case.

In Sec. 3, the classical and quantum SQM is formulated and the meson
wave functions are obtained. In going from classical SQM to the real
one,
we take into account quark spins, canonical quantization and
nonstring,
short-range, quark-antiquark interaction. All these effects are of
the
same order and all are necessary for consistency of the model. The
quark spins are introduced at the classical level with the help of
anticommuting variables obeying constraints [3]. We add a special
term to the  Lagrangian to ensure conservation of the spin
constraints, which renders the total SQM Lagrangian supersymmetric.
The canonical quantization implies finding out all the constraints
between canonical variables, and using a first form
method [4] to obtain the Poisson brackets of physical variables. As a
result, the meson wave function satisfies two Dirac equations and a
spectral condition, into which we introduce a nonstring, short-range,
contribution. In general, the spectral condition may contain
contributions
dependent on the meson spin. Since we believe that the long-range
contribution is given by the string term, then the additional
short-range
contribution can not increase with the meson spin. Experiment
suggests 
that the
short-range contribution does not depend on the spin or, for heavy
quarkonia,
has an additional, decreasing with the spin, term [1]. In this way we
have
phenomenological short-range parameters which depend on the type of
the
trajectory (i.e., on the space and charge-conjugation parity of the
wave
function) and on the quark masses. They obey the chiral symmetry
(then the
model obeys this symmetry) and, at present, are to be obtained from
experiment.  

In Sec. 4 and Appendix C, the spectral conditions for different meson 
wave functions are
compared with the experimental meson spectrum, the model parameters
are 
obtained and predictions of masses and other quantum numbers of new
mesons
are made. The results of SQM are  compared with that of a potential
quark model [5].

Knowing parameters of the model, we calculate in Sec. 5 the internal 
structure of mesons: the average values of string and quark energies,
and projections of quark spins and orbital momentum for polarized
mesons, as well as average total quark spin and orbital momentum
squared, and 
spin-orbit correlation.

Sec. 6 contains conclusions. Some mathematical and phenomenological
details 
are considered in Appendices A, B and C.

\section{Classical string physics}
The behaviour of a straight-line Nambu-Goto string, with or without
point spinless quarks at the string ends, follows from the Lagrangian
of the next Section. This behaviour can be described in terms of the 
point-particle relativistic mechanics if we take from the Lagrangian
three
properties of the string. Let the string be in its rest frame, where
the
string is at rest as a whole, i.e., its 4-momentum is $(m,\bf 0)$.
The specific string properties are the following.

I. The internal self-interaction string parameter $a$, called string
tension,
can be used as a "rest mass density" of the string.

II. The ends of an open string move with the velocity of light
perpendicular
to the string direction. The open string rotates in a plane around
its
center with an angular velocity
\be
\omega=2/d,
\ee
where $d$ is the string length.

III. Point quarks with current masses $m_i$, $i=1,2$ at the ends of 
a rotating string do not move along
the string. The string with quarks rotates in a plane.
Its angular velocity and position of the rotation center are
determined
by equality of the centrifugal force and the string-tension force
\be
\frac{m_i \omega^2 l_i}{\sqrt {1-\omega^2 l^2_i}}=
a\sqrt {1-\omega^2 l^2_i},
\ee
where $l_i$ is the distance between the rotation center and the $i$-th 
quark. For zero-mass quarks, Eq. (2) is equivalent to Eq. (1). For
heavy
quarks $\omega l_i \ll 1$, and Eq. (2) reduces to
\be
  m_i\omega^2 l_i=a.
\ee

We see that the main peculiarity of the string is that it always
rotates, and
can not be stopped. If the quarks at the string ends are heavy and
move slowly,
so that their velocities $v_i \rightarrow 0$, then, from Eq. (3), 
$l_i=v_i^2 m_i /a \rightarrow 0$,
and the string disappears. All points of the string can not be at
rest, and
the notion "rest mass density" is not applicable literally to the
string.
The property I above is a definition, following from the string
Lagrangian.
The string dynamics can not be reduced to the point-particle
dynamics,
although all other properties of the string can be obtained with its
help.

To make illustrative estimates, we shall use the experimental value
of $a$
\be
a=0.176~GeV^2\approx 1~GeV/fm.
\ee
It is a huge "mass density" on the macroscopic scale.

{\it Open string}. From I and II, the energy of an open string,
equal to its mass, is
\be
E_0=m=\int_{-d/2}^{d/2}{\frac{a dx}{\sqrt{1-\omega^2
x^2}}}=\frac{1}{2}\pi ad,
\ee
or the length of the string is proportional to its mass
\be
d=\frac{2}{\pi a} m.
\ee
The heavier  a light-quark meson, the bigger it is. The lightest
meson, the
pion, would have $d\approx 0.1~fm$.

>From II, the angular velocity of the string is inversely proportional
to its 
mass
\be
\omega=\frac{\pi a}{m}.
\ee
For the pion it would be $\omega\approx20~fm^{-1}\approx 10^{24}~Hz$.
We shall
see that the pion is not "the smallest top", but it is "the fastest
one".

In the same way we can calculate the angular momentum of the string
with
respect to its rotation center
\be
L=\int_{-d/2}^{d/2}{x~\frac{\omega x}{\sqrt{1-\omega^2 x^2}}\ a dx}=
\frac{\pi a}{2\omega^2},
\ee
or
\be
L=\frac{1}{2\pi a} m^2.
\ee
Both sides of this equation are observable. This is a well-known
linearly
(with respect to $m^2$) rising Regge trajectory.

For the pion Eq. (9) yields $L\approx 0.02$, a comfortably small
number.

{\it Heavy-quark mesons}. Let us introduce a meson mass excess
\be
m_E=m-m_1-m_2,
\ee
where $m$ is the meson mass and $m_1$ and $m_2$ are the current quark
masses,
and let us consider
\be
m_E/m_i\ll 1.
\ee 
Then the motion is nonrelativistic, and the energies of the string
and the 
quarks in Fig. 1b are
\be
E_0=ad,
\ee
\be
E_i=m_i+\frac{1}{2}al_i.
\ee
The last equation follows from Eq. (3). Summing all these equations,
we get
\be
d=\frac{2}{3a} m_E.
\ee

The contribution of the string energy to the meson mass is small, but
the
contribution to $m_E$ is not small,
\be
E_0/m_E=2/3=67\%,
\ee
and do not depend on the quark masses.

Eq. (14) reminds Eq. (6), where $m$ is replaced by $m_E$ and the
slope is
slightly bigger, to the extent that 3 is smaller than $\pi$.

We see that the string length in this case can be very small if $m_E$
is
small. Indeed, for the strange-quark current mass 0.22~GeV (Sec. 4),
the
diameter of the $\eta$-meson is smaller than that of the pion by
20\%. The
smallest particle is $\Upsilon$, the $b$-quark mass being 4.71~GeV
(Sec. 4).
The $\Upsilon$ diameter is 0.02~fm, 1/5 that of the pion.

On the contrary, the behaviour of the angular velocity of the
heavy-quark 
mesons is quite
different from the open-string case. From Eqs. (3) and (14) it is
easy to get
\be
\omega=\frac{a}{\sqrt{{\frac{2}{3}}\mu m_E}},
\ee
where $\mu$ is the reduced quark mass
\be
\mu=m_1 m_2/(m_1+m_2).
\ee
The angular velocity of $\Upsilon$ is 1/5 that of the pion.

The string angular momentum is negligible in this case and the total
angular
momentum is sum of the quark angular momenta
\be
L=\sum{m_i \omega l^2_i}=\frac{a^2}{\omega^3 \mu},
\ee
or
\be
L=\frac{1}{a}\Bigl(\frac{2}{3} m_E\Bigr)^{3/2} \mu^{1/2}.
\ee
This is also an observable Regge trajectory, nonlinear in this case,
but
determined by the same parameter $a$. 

{\it Asymmetric mesons}. Let one quark, with mass $m_1$, be heavy,
and the
other one be very light, i.e.,
\be
m_E/m_1\ll 1,~~~m_2/m_E\ll 1,
\ee
where
\be
m_E=m-m_1.
\ee
The string and quark energies are
\be
E_0=al_1+\frac{1}{2}\pi al_2,
\ee
\be
E_1=m_1+\frac{1}{2}al_1,
\ee
where, to a first approximation, $l_1$ can be neglected, and we
obtain
\be
d=\frac{2}{\pi a} m_E,
\ee
\be
E_0/m_E=1,
\ee
\be
\omega=\frac{\pi a}{2m_E},
\ee
\be
L=L_0=\frac{\pi a}{4\omega^2},
\ee
or
\be
L=\frac{1}{\pi a} m_E^2.
\ee
The diameter (24) has the same slope as that for the open string, Eq.
(6),
the string gives the main contribution to the meson mass excess $m_E$
and 
the Regge trajectory (28), as a function of $m_E^2$, has slope twice
as big 
as that for the open 
string, Eq. (9), although the corrections to
the first approximation,
which can be easily worked out, are not negligible in practice.

{\it General mesons}. For arbitrary quark masses
\be
m=E_0+\sum{E_i}=a \int_{-l_1}^{l_2}{\frac{dx}{\sqrt{1-\omega^2
x^2}}}+
\sum{\frac{m_i}{\sqrt{1-\omega^2 l^2_i}}},
\ee
\be
L=L_0+\sum{L_i}=a\omega\int_{-l_1}^{l_2}{\frac{x^2dx}{\sqrt{1-\omega^
2 x^2}}}+
\sum{\frac{m_i l_i^2\omega}{\sqrt{1-\omega^2 l_i^2}}},
\ee
where $l_i$ is given by Eq. (2).

Introducing
\be
l=1/\omega,
\ee
\be
l_i=\sqrt{l^2+m_i^2/(4a^2)}~ -m_i/(2a),
\ee
\be
G(l)=a\int\limits^{l_2}_{-l_1}\sqrt{l^2-x^2}~dx+\sum{m_i\sqrt{l^2-l^2
_i}}\,
\ee
we can rewrite Eqs. (29) and (30) in the form
\be
m=G_l(l)\equiv y\sum{(\arctan t_i + t_i^{-1})},
\ee
\be
L=K(l)\equiv\frac{1}{2a}(ym-\sum{m_i^2t_i}),
\ee
where index $l$ means derivative with respect to $l$, ~$y=al$, ~
$t_i=(al_i/m_i)^{1/2}$ and
\be
K(l)=lG_l(l)-G(l).
\ee
Eqs. (34) and (35) define a Regge trajectory as an implicit function
\be
L=K(l(m)),
\ee
where $l(m)$ is a solution of Eq. (34).

If the string moves as a whole with a velocity $\bf v$, its rotation
slows
down: the angular velocity acquires a factor $\sqrt{1-{\bf v}^2}$.
Its
length, in general, is not conserved. The length oscillates between
its
minimal (rest-frame)  value $d$, when the string is perpendicular to
the 
velocity,
and its maximal value $d/\sqrt{1-{\bf v}_{pl}^2}$, when the string is
parallel to the projection of the velocity on the rotation plane 
${\bf v}_{pl}$.

The classical description might be not only illustrative for $L\gg
1$. To make
the model realistic, we must quantize it and take into account quark
spins and
nonstring short-range interactions. This will be done in the next
Section.

\section{Quantum string physics}
We shall use Lorentz- and gauge-covariant variables. The
straight-line string
is a straight line in the 4-dimensional space-time
\be
X(\tau,\sigma)=r(\tau)+f(\tau,\sigma)q(\tau),
\ee
where $\tau$ and $\sigma$ are time-evolution and space-position
parameters,
respectively, $r$ is a 4-vector of a point on the straight line, $q$
is an
affine 4-vector of its direction, $f$ is a Lorentz scalar labelling
points on
the string, and $f_i=f(\tau,\sigma_i)$, $i=1,2$ correspond to the
string
ends. The covariant description introduces superfluous, from physical
viewpoint, variables, therefore, the string action must be invariant
with
respect to three $\tau$-dependent gauge transformations: shift of $r$
along
$q$, multiplication of $q$ by a function of $\tau$, and
reparametrization of 
$\tau$.
The Lagrangian must be invariant with respect to the first two
transformations
and have a property ${\cal L}(c\dot z)=c{\cal L}(\dot z)$, where
$\dot z$ is
every $\tau$-derivative and $c$ is a function of $\tau$. There is
only one 
string variable
which is Poincar\'e- and gauge-invariant
\be
l=\sqrt{\dot r^2_\perp}/b,
\ee
(not to consider higher $\tau$-derivatives), where $\dot r^\mu_\perp$
is
 the string velocity, perpendicular to the rotation plane,
\be
\dot r^\mu_\perp=(g^{\mu\nu}+n^\mu n^\nu -
\dot n^\mu \dot n^\nu/\dot n^2)\dot r_\nu,
\ee
\be
 n=q/\sqrt{-q^2},
\ee
and $b$ is an angular velocity of the string with respect to the
auxiliary
time $\tau$
\be
b=\sqrt{-\dot n^2}.
\ee
$b$ is gauge-dependent, but the condition
\be
b\not=0,
\ee
which we assume, is gauge-independent since $\tau$ is monotonous.
Then there
is a physically distinguished point on the string, the instantaneous
rotation
center, and we can label the points on the string, in a
gauge-invariant way, 
with respect to this center
\be
x=\sqrt{-q^2}\;f-\dot r\dot n/b^2.
\ee

The classical Lagrangian of a meson in SQM consists of three terms
\begin{equation}
{\cal L}={\cal L}_{str}+\sum{{\cal L}_i}+{\cal L}_{ss},
\end{equation}
the first one being the Nambu-Goto Lagrangian for a straight-line
string,
Eqs. (38)-(44), 
\be
{\cal L}_{str}=-ab\int\limits^{x_2}_{x_1}\sqrt{l^2-x^2}\;dx,
\ee
where $a$ is a string-tension parameter.

The second term in Eq. (45) is  sum of the Lagrangians for point
massive spinning quarks [BM] having velocities of the string ends 
\be
{\cal L}_i=-m_i \sqrt{\dot X^2_i}-\frac{i}{2}\xi_i^M\dot\xi_{iM}-i
\left (
\frac{\dot X_i\xi_i}{\sqrt{\dot X^2_i}}-\xi^5_i
\right )
b\lambda_i,
\ee
where $m_i$ is the quark current mass and $\xi^M_i$ and
$\lambda_i$ are quark-spin variables ($M=\mu,5$, and 
$g^{MN}=diag\{1,-1,-1,-1,-1\}$), which anticommute with each other
(and
commute with other variables, including spin variables of the other
quark).

The Lagrangian (47) contains spin-independent part
\be
{\cal L}_{i0} = -m_i \sqrt{\dot X_i^2},
\ee
spin-velocity term showing that that the spin variables $\xi_i^M$ are
canonically self-cojugate, and spin-constraint term, proportional to
a 
 Lagrange multiplier $\lambda_i$. The spin constraints must be
 conserved.
To ensure this conservation, we shall find out the third term in Eq.
(45)
(which  restore a supersymmetry of the total Lagrangian). Toward this
end,
let us first consider the spin-independent part of the Lagrangian
\be
{\cal L}_0 = {\cal L}_{str} + \sum{{\cal L}_{i0}}.
\ee
The quark velocity $\dot X_i$ must be perpendicular to the string
direction. This is a property of the minimal surface formed by
straight lines which follows from the Euler-Lagrange equations for
the full string under the assumption Eq. (38). The proof of this
property
is given in Appendix A.

Introducing orthonormal vectors
\be
v^0 = \dot r_\perp/(bl), ~~~ v^1 = \dot n/b,
\ee
we can write
\be
\dot X_i = b(lv^0 + x_i v^1).
\ee
The extremum condition for ${\cal L}_0$ with respect to $x_i$ yields
\be
x_i = (-1)^i l_i,
\ee
(Eq. (32)), and the Lagrangian takes the form
\be
{\cal L}_0 = -bG(l),
\ee
where $G(l)$ is given by Eq. (33). 

Let us rewrite this Lagrangian in the
phase-space. The   momenta conjugate to the coordinates $r$ and $q$
\be
p=-\partial {\cal L}/\partial\dot r,\;\;\;
\pi=-\partial {\cal L}/\partial \dot q.
\ee
are equal to
\be
p = G_l(l)v^0, 
\ee
\be
\pi = (-q^2)^{-1/2}((\dot r v^1/b)p + K(l)v^1),
\ee
where index $l$ stands for derivative with respect to $l$ and $K(l)$
is
given by Eq. (36).
The momentum $p$ is conserved due to  translation invariance. It is
the
total meson momentum, and the meson mass is
\be
m=\sqrt{p^2}.
\ee 
We shall use the notations
\be
n^0=p/m, ~~~~ \pi_p^{\mu}=(g^{\mu\nu}-n^{0\mu}n^{0\nu})\pi_{\nu}.
\ee
>From Eqs. (55) and (56)
\be
\pi_p=(-q^2)^{-1/2}Kv^1.
\ee
We see that the phase-space variables obey three constraints
\be
pq=0, ~~~~ \pi q=0,
\ee
\be
m=G_l(l),
\ee
\be
\sqrt{q^2 \pi_p^2}=K(l).
\ee
The third constraint is given by two Eqs. (61) and (62): we must
solve one
of them (e.g., the first one) to find out $l$ as a function of $m$,
and put
this solution into the second equation. The l.-h.s. of Eq. (62) with
constraints Eqs. (60) is (orbital) angular momentum of our system
\be
\sqrt{q^2\pi_p^2}=\sqrt{-L^2},
\ee
\be
L_\mu=\epsilon_{\mu\nu\rho\sigma} p^\nu L^{\rho\sigma}/2m,~~~
L^{\mu\nu}=r^{[\mu}p^{\nu]}+q^{[\mu}\pi^{\nu]}.
\ee
The canonical Hamiltonian of a $\tau$-reparametrization-invariant
system
is zero, and the Hamiltonian of our system is a linear combination of
the
constraint functions [6]. We can rewrite the Lagrangian (49), (53) in the
form
\ba
&{\cal L}_0&=-(1/2)(p\dot r-r\dot p)-(1/2)(\pi\dot
q-q\dot\pi)-\nonumber\\
&-& c\left(\sqrt{-L^2}-K(l(m))\right)-c_1 pq-c_2\pi q.
\ea
where $c, c_1$ and $c_2$ are arbitrary ($c=-b$) and $l(m)$ is given
by 
Eq. (61).

>From Eq. (48), we get the quark momenta
\be
p_i=m_i \dot X_i/\sqrt{\dot X_i^2},
\ee
which, from Eqs. (51), (52), (55), and (59), are equal to
\be
p_i=(ln^0+(-1)^il_i n^1)m_i/\sqrt{l^2-l_i^2},
\ee
where $l=l(m)$ is given by Eq. (61) and
\be
n^1=\pi_p/\sqrt{-\pi_p^2}.
\ee

Now it is not difficult to introduce the quark spins in a consistent
way.
We add to the r.-h.s of Eq. (65) the spin-velocity term and the 
spin-constraint term expressed through the quark momenta (67), and,
to
ensure the spin-constraint conservation, we replace the orbital
angular
momentum $L_\mu$ by total angular momentum 
\be
J_\mu=\epsilon_{\mu\nu\rho\sigma} p^\nu M^{\rho\sigma}/2m,
\ee
\be
M^{\mu\nu}=r^{[\mu}p^{\nu]}+q^{[\mu}\pi^{\nu]}-i\sum
\xi^\mu_i\xi^\nu_i.
\ee
As a result, we obtain the SQM Lagrangian
\ba
&{\cal L}&=-(1/2)(p\dot r-r\dot p)-(1/2)(\pi\dot q-q\dot\pi)-(i/2)
\sum{\xi^M_i\dot\xi_{iM}}-\nonumber\\
&-&i\sum{(p_i\xi_i-m_i\xi^5_i)\lambda_i}-
c\left(\sqrt{-J^2}-K(l(m))\right)-
c_1 pq-c_2\pi q.
\ea
One can express this Lagrangian through the configuration space
variables
 by means of the inverse Legendre transformation. We shall not use
 the
configuration-space form of the Lagrangian. For the sake of
compliteness,
it is given in Appendix B with an outline of its derivation.

Besides quark-spin terms, a nonstring short-range interaction must
enter
into the meson Lagrangian. This interaction cannot be fully described
at the classical level and will be taken into account in the quantum
equations.

The first line of Eq. (71) corresponds to the first differential form
of our
system which determines the Poisson brackets of the canonical
variables [4].
Namely, if we denote the variables by $y_n$ and the first form by
$(1/2)\omega_{mn}y_m\dot y_n$, then the Poisson brackets are
$\{y_m,\;y_n\}=\omega^{mn}$, where $\omega^{mn}$ is inverse of
$\omega_{mn}$.
For instance, from Eq. (71),
\be
\{\xi^M\;,\;\xi^N\}=ig^{MN}.
\ee
The other brackets have the usual canonical form. In particular, the
total spin
$J^{\mu}$ has zero Poisson brackets with Lorentz scalars, therefore,
the spin
constraint functions have zero brackets with the Hamiltonian and are
conserved.
This justifies the choice of ${\cal L}_{ss}$ in the Lagrangian. The
gauge
constraint functions are in involution with respect to the Poisson
brackets,
due to properties of the gauge transformations, and are also
conserved.

The second line of Eq. (71) is minus Hamiltonian. We can exclude the
constants
$c_1$ and $c_2$ by choosing the gauge conditions
\be
p\pi=0,\;\;\; \pi^2=-1.
\ee
Their conservation yields $c_1=c_2=0$, and we must solve Eqs. (73)
together
with the corresponding constraints (60). Introducing  four
orthonormal vectors 
$e_\alpha,\;\; \alpha=0,1,2,3$
\be
e_0=p/m,\;\;\; e_\alpha e_\beta=g_{\alpha\beta},
\ee
we can write the solution in the form ($a,b,c=1,2,3$)
\be
\pi=k^{(a)}e_a,\;\;\;q=\epsilon_{abc}k^{(a)}L^{(b)}e_c,\;\;\;L=L^{(a)
}e_a.
\ee
We shall use space-vector notations for the set $\{k^{(a)}\}$ and
similar sets 
\be
\{k^{(a)}\}=\vec k.
\ee
>From Eqs. (73) and (64)
\be
\vec k^2=1,\;\;\;\vec k\vec L=0.
\ee
Using expansions
\be
J=J^{(a)}e_a,\;\;\;\xi_i=\xi_i^{(\alpha)}e_{\alpha}\;,\;\;
\xi^{(5)}_i=\xi^5_i,
\ee
we can rewrite the Lagrangian (71) in the form (up to a total
derivative)
\ba
&{\cal L}&=-p\dot z-[\vec k\times \vec L]
\dot{\vec k} -\frac{i}{2}\sum{\xi_{i(M)}
\dot\xi^{(M)}_i} -  \nonumber\\
&-& c(\sqrt{\vec J^2}-K)+
\sum{ (p^{(\alpha)}_i\xi_{i(\alpha)}-m_i\xi^5_i)\lambda_i},
\ea
where the new string coordinate is
\be
z^\mu=r^\mu+\frac{1}{2}\epsilon_{abc} 
e^{\nu}_a\frac{\partial e_{b\nu}}{\partial p_{\mu}}J^{(c)}+
\frac{i}{m}\sum{\xi^{(0)}_i
\xi^{(a)}_i e^\mu_a},
\ee
 and, in 4-vector notations,
\be
p^{(\alpha)}_i=(l,~~(-1)^i l_i\vec k)m_i/\sqrt{l^2-l_i^2}.
\ee
The variables $\vec k$ and $\vec L$ are not independent, but using,
e.g.,
spherical angles to solve Eqs. (77), one can easily obtain from the
Lagrangian (79) the following nonzero Poisson brackets
\be
\{p^\mu,z^\nu\}=g^{\mu\nu}
\ee
\be
\{L^{(a)},L^{(b)}\}=\epsilon_{abc}L^{(c)},\;\;\;
\{L^{(a)},k^{(b)}\}=\epsilon_{abc}k^{(c)}
\ee
\be
\{\xi_i^{(M)},\xi_i^{(N)}\}=ig^{MN}.
\ee

The Hamilton equations of motion can be easily solved. The solution
for 
spinless quarks was described in Sec. 2.

The quantization of our system is now straightforward. We replace
$p$, $ z$, $\vec L$, $\vec k$ and $\xi^{(M)}_i$ by operators and
their
Poisson brackets  by commutators or anticommutators for $\xi'$s 
(multiplied by $-i$), e.g.,
\be
[\xi^{(M)}_i, \xi^{(N)}_i]_+=-g^{MN},\;\;\;[\xi^{(M)}_1,
\xi^{(N)}_2]_-=0.
\ee
Assuming the second quark to be an antiquark, we take the following
solution
of these equations
\be
\xi^{(\mu)}_1=\frac{1}{\sqrt{2}}\gamma^5\gamma^\mu\otimes I,\;\;\;
\xi^{(5)}_1=\frac{1}{\sqrt{2}}\gamma^5\otimes I
\ee
\be
\xi^{(\mu)}_2=\xi^{(\mu)c}_1=I\otimes
\frac{1}{\sqrt{2}}\gamma^{5c}\gamma^{\mu c},\;\;\;
\xi^{(5)}_2=\xi^{(5)c}_1=I\otimes
\frac{1}{\sqrt{2}}\gamma^{5c},
\ee
where $\gamma^c=C\gamma C^{-1}$ and $C$ is the charge-conjugation
matrix.

The constraint functions become
operators annihilating the wave function. In the representation where
$p$ and $\vec k$ are diagonal the internal part of the wave function
$\delta (p-p')\Psi_{\alpha\beta}(\vec k)$ satisfies the equations
\be
(\hat p_1-m_1)\Psi=0\;,
\ee
\be
\Psi (\hat p_2+m_2)=0\;,
\ee
\be
(\sqrt{\vec J^2} - K - \sum^{4}_{n=1}a_nP_n)\Psi=0.
\ee
In the third equation a new term has been introduced to account for
the short-range nonstring interaction. In this term, $a_n$ can depend
on $J$, and $P_n$ are four independent  operators commuting with the
Dirac operators in the first and second equations (for fixed
$J\not=0$ there are four independent states of two particles with
spin 1/2). 

Since $a_n$ describes a short-range interaction, it cannot increase
with $J$. For the majority of mesons we can take $a_n$ as a constant
independent of $J$. Only heavy quarkonia demand more complicated
$a_n$, containing a decreasing with $J$ contribution [1].

The choice of $P_n$ is connected with the choice of meson states at
fixed $J$ and will be discussed after solving Eqs. (88) and (89).

The solution of the Dirac equations (88) and (89) is a $4\times 4$
matrix
\be
\Psi=\frac{1}{\sqrt{(1+b^2_1)(1+b^2_2)}}
\left (
\begin{array}{cc}
-b_2\chi\vec\sigma\vec k & \chi\\
b_1b_2\vec\sigma\vec k\chi \vec\sigma\vec k & -b_1\vec\sigma\vec
k\chi
\end{array}
\right )
\ee
where
\be
b_i=\frac{l_i}{l+\sqrt{m_il_i/a}},\;\;\;
b_i\rightarrow
\left\{
\begin{array}{cc}
1, & m_i\to 0\\
0, & m_i\to \infty
\end{array}
\right.
\ee
and $\chi$ is an arbitrary normalized $2\times 2$ matrix.

We shall take
\be
\chi=\chi_{jMlS},
\ee
which are eigenfunctions of $\vec j^2, j^{(3)}, \vec L^2$ and $\vec
s^2$ with eigenvalues $j(j+1), M, l(l+1)$ and $S(S+1)$, respectively,
where $\vec j=\vec L+\vec s$ and $\vec s$ is a 2-dimensional quark
spin
\be
\vec s=\frac{1}{2}\vec\sigma\otimes
1+\frac{1}{2}1\otimes\vec\sigma^c,\;\;\;
\vec\sigma^c=\sigma_2\vec\sigma\sigma_2=-\vec\sigma^*
\ee
and 1 stands for the $2\times 2$ unit matrix. These functions can be
easily constructed with the help of Clebsch-Gordon coefficients.

The corresponding functions $\Psi$, denoted by $\Psi_{jMlS}$, are
eigenfunctions of $\vec J^2$ and $J^{(3)}$ with eigenvalues $j(j+1)$
and $M$, respectively, and eigenfunctions of space and
charge-conjugation 
parities 
\be
P\Psi_{jMlS}=-(-1)^l\Psi_{jMlS},\;\;\;
\Psi^c_{jMlS}=(-1)^{l+S}\Psi_{jMlS}.
\ee
The parity transformations are defined by
\be
P\Psi(\vec k)=\gamma^0\Psi (-\vec k)\gamma^0,\;\;\;
\Psi^c(\vec k)=C\Psi^T(-\vec k)C^T,
\ee
where $C$ is the charge-conjugation matrix. The Dirac equations (88),
(89) are charge-conjugation-invariant for $m_1=m_2$.

We shall assume that mesons in the states $\Psi_{jM,j-1,1}$ and
$\Psi_{jM,j+1,1}$ do not mix.

This means that mesons with definite $C$ or $G$-parity are described
by the wave functions $\Psi_{jMlS}\equiv \Psi_n$, and the operators
$P_n$ in Eq.(90) are projection operators
\be
P_n\Psi_m=\delta_{mn}\Psi_n.
\ee
The index $n$ takes four (two) values for fixed $j\not= 0(j=0):\;\;
n=0$ for $l=j,\;\; S=0;\;\; n=1$ for $l=j,\;\; S=1;$ and $n=\pm$ for
$l=j\pm 1,\;\; S=1$. Eq. (90) for these states takes the form
\be
\sqrt{j(j+1)}=K+a_n
\ee
which is called the spectral condition. Here $K$ is a  function of
$m,
m_1, m_2$ and $a$, given by Eqs. (36), (33), and (32).

As mentioned before, $a_n$ can be taken independent of $j$ for all
mesons except $c\bar c$- and $b\bar b$-mesons in the states with
$n=-$, for which
\be
a_-=A+\left (\frac{8m_1}{m(2j+1)^2}\right )^2B\;,
\ee
where $A$ and $B$ do not depend on $j$.

For strange, charmed and bottom mesons the states $\Psi_0$ and
$\Psi_1$ can mix for $j>0$, so that the mesons are described by
\ba
\Phi_0&=&\cos\alpha\;\Psi_0+\sin\alpha\;\Psi_1\nonumber \\[-0.2cm]
\\[-0.2cm]
\Phi_1&=&-\sin\alpha\;\Psi_0+\cos\alpha\;\Psi_1,\nonumber
\ea
and in the spectral condition (98) $a_0$ and $a_1$ are replaced by
$a_0-d$ and $a_1+d$, respectively, where $d$ is a mixing parameter
\be
d=-b\tan \alpha=\frac{1}{2}(a_0-a_1\pm
\sqrt{(a_0-a_a)^2+4b^2})
\ee
and upper (lower) sign corresponds to $a_0-a_1<0(>0)$.



\section{Meson mass spectrum and model parameters}
First comparison of the spectral condition (98) with
experiment was made in Ref. [1,2]. Here we compare with more recent
data
available [7] and make predictions for mesons with spin up to 7.

The light quark current masses give very small contribution to the
spectral condition and can not be determined from this condition. So
we use the linear chiral $SU_3$ model relations [8]
\be
m_u/m_d=0.55,\;\;\;
m_s/m_d=20.1
\ee
to express them through the strange quark mass which is determined
from
comparison with experiment.

The best known meson Regge trajectories are described by the wave
functions 
$\Psi_-$ and have $P=C=(-1)^j$ and $j_{\min}=1$.
The parameters $a_-$ or $A$ in Eq. (99) depend very weakly on the
quark 
masses: 
$a_-(d\bar u)=0.88$,~~$ a_-(c\bar u)=0.90$,~~$A(c\bar c)=0.90$,~~  
$a_-(b\bar u)=0.77$, and $A(b\bar b)=0.77$. This means that the
short-range
contributions for the strange quarks in these states are the same as
for
the light quarks: $a_-(s\bar u)=a_-(s\bar s)=a_-(d\bar u)$,
$a_-(c\bar s)=a_-(c\bar u)$, and $a_-(b\bar s)=a_-(b\bar u)$. 

These trajectories allow one to determine the main parameters of the
model~[1,2]
$$
a=0.176\pm 0.002\;\;\mbox{GeV}^2
$$
\be
m_s=224\pm 7,\;\;
m_c=1440\pm 10,\;\; m_b=4715\pm 20
\ee
$$
m_u=6.2\pm 0.2,\;\;
m_d=11.1\pm 0.4 
$$ 
(masses in MeV), and the short-range parameters, Table IV in Appendix
C.

A simpler procedure, when one drops the second term in Eq. (99), uses
$a_-$
independent of the quark masses and applies the
minimum-$\chi^2$-method [2],
gives the same results for the main parameters (103) with good
$\chi^2$.

 For the light- and strange-quark 
mesons the trajectories are practically linear.

For the light-heavy-quark mesons the trajectories are not linear but
can be made practically linear by replacing the argument $m^2$ by
$(m-m_h)^2$ where $m_h$ is the heavy quark mass.

In the limit
\be
\frac{2(m-m_h)}{\pi m_h}\ll 1,\;\;\;
\frac{\pi m_l}{2(m-m_h)}\ll 1,
\ee
where $m_l$ is the light-quark mass, they must be linear with the
slope twice as big as for the light-quark mesons.  The
trajectories are practically linear up to $j=5$ with bigger effective
slopes, but the first condition (104) for the limit slope is not
fulfilled.

The trajectories for the heavy quarkonia are essentially nonlinear.

Table IV in Appendix C represents a detailed comparison of the model
with
experiment and contains predictions for new mesons and comparison
with a
potential model predictions [5]. The $B_J^*(5732)$-meson, found in
ALEPH, 
DELPHI and OPAL experiments at CERN
(see page 574 of Ref. [7] for the References) agrees with SQM much
better than
with PQM,  Table IV. 

The prediction for the $b\bar c, 1^-$-meson
is made under the simplest assumption $a_-(b\bar c)=a_-(b\bar u)$.
The
other assumptions: $a_-(b\bar c)=a_-(b\bar b)$ or $a_-(c\bar c)$
reduce its 
mass
by a 100~ MeV. The higher-spin $b\bar c$-mesons practically do not
depend on
this assumption.

The trajectories for $\Psi_0$  states, $C=-P=(-1)^j$ and $j_{\min}=0$
are always nonlinear near $j=0$ due to the square root in the
spectral condition (98).

Only one parameter, the short-range contribution
$a_0$, is unknown for each of these trajectories.
It is determined from the mass of corresponding spin-0 meson.

 We see that the nonlinear trajectory (98) with
the universal slope describes quite well the three mesons $\pi, b_1$
and $\pi_2$.

The constant $a_0$ for the light-quark mesons is small. According to
the linear chiral $SU_3$ model [8],  it must be proportional to the
light-quark masses $m_l$. This property does not contradict the
spectral 
condition (98) where $K$ is proportional to $m_l^{3/2}$.

There is not enough data to analyse $\Psi_1$ states, $P=C=-(-1)^j$
and
$j_{\min}=1$. So we assume that
\be
a_1=a_0
\ee
for all mesons except $s\bar u$- and $s\bar d$-mesons for which
mixing is
important. Experiment confirms this assumption for known mesons
composed by
light, $s\bar s$- and $c\bar c$-quarks. If Eq. (105) is fulfilled
also
for the $b\bar b$-mesons then using the known mass of $\chi_{b1}(1P)$
we can estimate the mass of a pseudoscalar $b\bar b$-meson 
$\eta_b(1S), 0^{-+}$ to be 9.50~GeV which is bigger than the mass of
$\Upsilon (1S)$.

Linearly extrapolating between $a_0(b\bar u)=-0.55$ and 
$a_0(b\bar b)=-0,091$, obtained from Eq. (105), we can find 
$a_0(b\bar c)=-0.41$ and estimate the mass of a pseudoscalar
$B_c$-meson to be
6.40~GeV, which is 0.13~GeV higher than in the potential model~[5].

We take into account mixing of $\Psi_0$ and $\Psi_1$ states only for
strange mesons. The mixed states $\Phi_0$ and $\Phi_1$ 
(100) correspond to a  mixing angle $36^\circ$. The
detailed
comparison with experiment and predictions for $\Psi_{0,1}$
($\Phi_{0,1}$)
states are given in Table V in Appendix C.

The behaviour of the trajectories for $\Psi_+$ states, $P=C=(-1)^j$
and
$j_{\min}=0$, is similar to that for $\Psi_0$. 

The $X(1920), ?^{??}$-meson, found in GAMC and VES experiments at
IHEP, 
Protvino, agrees quite well with SQM predictions and may be a $2^{++}$ 
trajectory partner of $a_0(980)$.

The strange mesons $K^*_0(1430), 0^+$ and $K^*(1680), 1^-$ are not
described by the same wave function $\Psi_+$ (with different $j$). It
seems probable that a new strange $1^-$ meson exists with mass
1900~MeV which is a partner of $K^*_0(1430)$, see Table VI in
Appendix C.
On the other hand, the $K^*(1680)$-mass, $1717\pm 17$ MeV, is only
half
of its width, $322\pm 110$ Mev, lower than the SQM value $1910$ Mev.

We can tentatively conclude that the SQM descrides masses and other
quantum
numbers of about 2/3 of established mesons, the rest being daughter,
gluball, 
or exotic states. The agreement with experiment for the former mesons
is in
general slightly better than that for the PQM. It seems important to
continue
systematic experimental study of meson mass spectrum where both
models give
different new predictions.

In conclusion of this Section, let us compare the description of heavy
quarkonia in SQM and PQM. In the frame where the meson is at rest as a
whole and the evolution parameter is the laboratory time, the SQM 
Hamiltonian is $m(J)$ where $m$ is the solution of the spectral condition.
Let us take $\Upsilon$ for definiteness and neglect the b-quark kinetic
energy. Then the Hamiltonian is the sum of three terms
$$H=2m_b+H_{string}+H_{short-range}$$
Considering the spectral condition without short-range contribution and
using the current b-quark mass from Table 1C, 
it is not difficult to estimate the last term, so that (in MeV)
$$H=9430+350-320$$
(Note that in Table 1 in the next Section the parameter $E_0$ comprises 
the string and the short-range contributions since it corresponds
to $x$ with account of the short range contribution in the spectral
condition.)

We see that the string contribution is comparable with the short-range
contribution. The confining role of the string for $\Upsilon$ is small:
the string-tension energy is $a(l_1 + l_2)$=$2x^2/m_b$=20~MeV; but the
kinetic energy of the string is not negligible.

The PQM Hamiltonian also contains three terms
$$H=2M_b+H_{conf}+H_{short-range},$$
where $M_b$ is the constituent quark mass which is bigger than the current
mass. In Ref. [5] $M_b$=4977~MeV, $H_{conf}=c+br$= --70~MeV for $\Upsilon$,
so that in this case
$$H=9950-70-420.$$
We see that the confinement potential is also small, and we need string    
to use current quarks instead of constituent ones.

\section{Internal structure of composite mesons}
The model allows one to calculate quark velocities and energies and
string energy in mesons at rest, Eqs. (61) and (81):
\be
v_i=l_i/l,\;\;E_i=l(am_i/l_i)^{1/2},\;\;E_0=al\sum{\arcsin v_i},
\ee
where $l_i$ is given by Eq. (32) and $l$ is a solution of Eq. (61),
$l=x/a$ and $x$ is given for each meson in Tables in Appendix C. The
results
for some mesons are collected in Table I.

\vspace*{0.4cm}
TABLE I.Energy distribution inside mesons at rest. $v_i(E_i)$ is
velocity in $c$ (energy in MeV) of the $i$-th quark, $E_0$ is energy
of the gluon string in MeV and $m_E=m-m_1-m_2$.

\begin{center}
\begin{tabular}{r|c|c|c|c|c|c|c|}
Particle, & & & & & &\\
quark content & $v_1$ & $v_2$ &  $E_1$ &  $E_2$ & $E_0$ &
$E_0/m,\%$ &
$E_0/m_E,\%$ \\
\hline
$\rho^+, d\bar u$ & 0,98 & 0,99 & 53 & 39 & 679 & 88 & 90 \\
$\pi^+, d\bar u$ & 0,88 & 0,93 & 23 & 16 & 99 & 72& 82 \\
$B^+, b\bar u$ & 0,07 & 0,99 & 4727 & 46 & 507 & 9.6 & 91 \\
$J/\psi(1S), c\bar c$ & 0,22 & 0,22 & 1476 & 1476 & 146 & 4.7 & 67 \\
$\Upsilon(1S), b\bar b$ & 0,05 & 0,05 & 4720 & 4720 & 22 & 0.2 & 67
\\
$\chi_{b2}(1P), b\bar b$ & 0,18 & 0,18 & 4795 & 4795 & 324 & 3.3 &
67\\
\hline
\end{tabular}
\end{center}

We see that the light quarks are
relativistic and give noticeable contributions to the meson masses.
 The main contribution to the mass ``excess'' of mesons
$m_E=m-m_1-m_2$ is given by the gluon string.

Let us consider spin structure of mesons, i.e., average values of
internal 
angular momentum variables. The SQM allows one to calculate the spin
structure of each
meson on leading trajectories. The result depends on spin, parities
and
mass of the meson, string tension and current masses of quarks
composing the meson. For instance, an average value of the third
projection 
of the i-th-quark spin is given by
\be
\overline{S^{(3)}_i}=(\Psi, S^{(3)}_i\Psi)=
\int Sp\Psi^+S^{(3)}_i\Psi d\vec k,
\ee
where
\be
\vec S_1=\frac{1}{2}\vec\Sigma\otimes I\;,\;\;\;
\vec S_2=\frac{1}{2}I\otimes \vec\Sigma^c,\;\;\;
\ee
\be
\vec\Sigma=
\left (
\begin{array}{cc}
\vec\sigma & o\\
0 & \vec\sigma
\end{array}
\right ),\;\;\;
\vec\Sigma^c=
\left (
\begin{array}{cc}
\sigma_2\vec\sigma\sigma_2 & 0\\
0 & \sigma_2\vec\sigma\sigma_2
\end{array}
\right )
=-\vec\Sigma^*.
\ee
Introduce the notations
\be
c=2(b_1^2+b_2^2)N,\;\;c_1=4b_1^2b_2^2N,\;\;c_2=2(b_2^2-b_1^2)N,
\ee
\be
N=1/((1+b_1^2)(1+b_2^2))
\ee
where $b_i$ is given by Eq. (92). In the nonrelativistic limit
($v_i=0$,
$m_1+m_2=m$) all $c$'s vanish. In the ultrarelativistic limit
($v_i=1$,
$m_i=0$) $c=c_1=1$ and $c_2=0$. Then,
for a polarized meson  with $J^{(3)}=M$, we obtain
\be
\Psi_0=\Psi_{jMj0},\;\;\;
\overline{S^{(3)}_i}=0,
\ee
\be
\Psi_1=\Psi_{jMj1},\;\;\;
\overline{S^{(3)}_i}=\frac{M}{2j(j+1)}\;,
\ee
\be
\Psi_-=\Psi_{jM,j-1,1},\;\;
\overline{S^{(3)}_i}=\frac{M}{2}\left (
\frac{1}{j}
-\frac{1}{2j+1}
(c+c_1+(-1)^ic_2)\right )\;,
\ee
\be
\Psi_+=\Psi_{jM,j+1,1},\;\;\;
\overline{S^{(3)}_i}=\frac{M}{2}\left (
-\frac{1}{j+1}+\frac{1}{2j+1}(c+c_1+(-1)^ic_2)\right )\;.
\ee
\be
\overline{S^{(3)}}=\sum{\overline{S^{(3)}_i}},\;\;\;
\overline{L^{(3)}}=M-\overline{S^{(3)}}.
\ee
In the same way, for squared quantities, we have
\be
(\Psi_0,\vec S ^2\Psi_0)=c,\;\;(\Psi_1,\vec S ^2\Psi_1)=2,
\ee
\be
(\Psi_-,\vec S ^2\Psi_-)=2-\frac{j}{2j+1}c,
\ee
\be
(\Psi_+,\vec S^2\Psi_+)=2-\frac{j+1}{2j+1}c.
\ee
\be
(\Psi_0, \vec L ^2\Psi_0)=j(j+1)+c,\;\;(\Psi_1,\vec L
^2\Psi_1)=j(j+1),
\ee
\be
(\Psi_-,\vec L ^2\Psi_-)=j\left(
j-1+c+\frac{2j+2}{2j+1}c_1\right),
\ee
\be
(\Psi_+,\vec L ^2\Psi_+)=(j+1)\left(
j+2-c-\frac{2j}{2j+1}c_1\right).
\ee
\be
\overline{\vec L\vec S}=(1/2)\left(
j(j+1)-\overline{\vec L ^2}-\overline{\vec S ^2}\right).
\ee
The spin structure of some mesons is represented in Tables~II~and~III.

\vspace*{0.4cm}
TABLE II. Spin structure of some mesons. Average values of internal
angular momentum variables are shown for polarized mesons with
$J^{(3)}=M$. nr--- nonrelativistic limit 
, r --- real case and ur --- ultrarelativistic limit
.

\begin{center}
\tabcolsep 4pt
\begin{tabular}{r|c|c|c|c|c|c|c|c|c|c|c|c|}
 & \multicolumn{3}{c}{$\overline{S^{(3)}_1}/M$}  \vrule & 
\multicolumn{3}{c}{$\overline{S^{(2)}_2}/M$}\vrule &
\multicolumn{3}{c}{$\overline{S^{(3)}}/M$}\vrule &
\multicolumn{3}{c}{$\overline{L^{(3)}}/M$}\\
  \cline{2-13} 
 & nr & r & ur & nr & r & ur &
nr & r & ur & nr & r & ur\\
\hline
$\rho^+, u\bar d$ & & 0,22 & & & 0,23 & & & 0,45 & & & 0,55 &
\\[-0.2cm]
    & 1/2 & & 1/6 & 1/2 & & 
1/6 & 1 & & 1/3 & 0 & & 2/3 \\[-0.2cm]
$K^{*+}, u\bar s$ & & 0,22 & & & 0,42 & & & 0,63 & & & 0,37 & \\ 
\hline
\end{tabular}
\end{center}

\vspace*{0.4cm}
TABLE III. Continuation of Table~II. Average values do not depend on the
meson polarization.

\begin{center}
\begin{tabular}{r|c|c|c|c|c|c|c|c|c|}
 & \multicolumn{3}{c}{$\overline{\vec{S}^2}$}  \vrule & 
\multicolumn{3}{c}{$\overline{\vec{L}^2}$}\vrule &
\multicolumn{3}{c}{$\overline{\vec{S}\vec{L}}$}\\
  \cline{2-10} 
 & nr & r & ur & nr & r & ur &
nr & r & ur \\
\hline
$\pi^+, u\bar d$ & & 0,83 & & & 0,83 & &  & -0,83 & \\[-0.2cm]
 & 0 &  & 1 & 0 &  & 1 & 0  &  & -1 \\[-0.2cm]
$K^+, u\bar s$ &  & 0,79 &  &  & 0,79 & & & -0.79 &  \\
\hline
$\rho^+, u\bar d$ & & 1,68 & & & 1,87 & & & -0,77 &   \\[-0.2cm]
 & 2 &  & 5/3 &0  &  & 7/3 &0  & & -1 \\[-0.2cm]
$K^{*+}, u\bar s$ &  & 1,70 &  &  & 1,17& &  &-0,43 &  \\
\hline
\end{tabular}
\end{center}

We see that the spin structure of light-quark mesons (114) is
essentially
different from the nonrelativistic case: the average
quark spin projections are twice as small. The spin structure of
$\rho$-meson
in SQM is similar  to the nucleon spin structure measured
in experiment and different from the nonrelativistic quark model
predictions.

The spin structure is also different from the ultrarelativistic case,
when the light-quark current masses are neglected. Unlike the
spectral condition, the spin structure is sensitive to the
light-quark current masses. Measurement of the spin structure allows
one to estimate the light-quark current masses from experiment.

We see also that the flavour $SU_3$ is badly broken in the spin
structure of spinning mesons. The average value of the $\bar s$ spin
projection in $K^*$ is 80\% bigger than the $\bar d$ spin projection
in $\rho$.

\section{Conclusions}
The gluon string in SQM can account for the quark confinement in
mesons.

The string comprises two mechanisms of potential quark models (PQM):
confinement potential and constituent quark masses.

Systematic experimental study of meson spectroscopy is important to
check the SQM predictions in comparison with the PQM predictions. 

Spin structure of light-quark vector mesons in SQM is different from
the
nonrelativistic quark model (NQM): for the average light-quark spin
projections $\bar S_{SQM}\cong \frac{1}{2}\bar S_{NQM}$.

The flavour $SU_3$ is badly broken in the spin structure in SQM: for
$s$ and $d$ quarks $\bar S_s\cong 2\bar S_d$.

Experimental study of the spin structure may eventually provide
experimental estimation of the light-quark current masses.
\\

The author is sincerely grateful to Professor A. D. Krisch for the
kind
hospitality at the Spin Physics Center of the University of Michigan.

\vspace*{0.4cm}
\begin{center}
{\large\bf Appendix A: Lagrangian for a straight-line string with
massive spinless quarks at the ends}\\
\end{center}
\vspace*{0.2cm}

This Lagrangian must give equations of motion which follow from the
full-string
Lagrangian with quarks at the ends, $i=1$ or $2$
\be
\left(\partial L/\partial \dot X\right)^\cdot +
\left(\partial L/\partial X'\right)'=0,
\ee
\be
(-1)^i\left((\partial L(\sigma_i)/\partial\dot X)\dot\sigma_i - 
\partial L(\sigma_i)/\partial X'\right)+
\left(\partial L_i/\partial \dot X_i\right)^\cdot=0,
\ee
where $L (L_i)$ is the Nambu-Goto (the $i$-th-quark) Lagrangian, and
dot 
(prime) stands
for derivative with respect to $\tau$ ($\sigma$). 
For a straight-line string in the notations of Sec. 3
\be
X(\tau,\sigma)=r(\tau)+(x(\tau, \sigma)+z(\tau))n(\tau),
\ee
\be
z=\dot rv^1/b,
\ee
\be
w=\dot x+\dot z-\dot r n,
\ee
we can rewrite Eq. (124) in the form
\be
\left(x'(lv^0+xv^1)/s\right)^\cdot-\left(w(lv^0+xv^1)/s+bsn\right)'=0
,
\ee
\be
s=\sqrt{l^2-x^2}.
\ee
Using four orthonormal vectors $v^0$, $v^1$ (Eqs. (50)), $n$ (Eq.
(41)) and
\be
v^2_\mu=\epsilon_{\mu\nu\rho\sigma}v^{0\nu}
v^{3\rho}v^{1\sigma},\;\;\;
v^3=n,
\ee
\be
v^av^b=g^{ab},\;\;\;a,b=0,1,2,3,
\ee
we can expand the l.-h.s. of Eq. (129) with respect to these vectors
and get
three equations (the fourth one, corresponding to the $n$-component,
turns out
to be an identity)
\be
\left(x'l/s\right)^\cdot+\alpha x'x/s-\left(wl/s\right)'=0,
\ee
\be
\left(x'x/s\right)^\cdot+\alpha x'l/s-\left(wx/s\right)'=0,
\ee
\be
\beta l+\gamma x=0,
\ee
where
\be
\alpha=-\dot v^0 v^1,\;\;\;\beta=-\dot v^0 v^2,\;\;\;\gamma=-\dot v^1
v^2.
\ee
Since $x$ is the only function which depends on $\sigma$, we get from 
Eq. (135)
\be
\beta=\gamma=0.
\ee
Eqs. (133) and (134) coincide
\be
\dot lx-\alpha(l^2-x^2)+(\dot z-\dot rn)l=0
\ee
and give
\be
\dot l=\alpha=\dot z-\dot rn=0.
\ee
So, Eq. (124) for the straight-line string is equivalent to
\be
\dot l=0,\;\;\;\dot v^0=0,\;\;\;\dot v^1=-bn,\;\;\;\dot z-\dot rn=0.
\ee

Let us consider Eq. (125)
\be
(-1)^i\left(\dot x_i(lv^0+x_iv^1)/s+bs_in\right)+
m_i\left(\dot X_i/\sqrt{\dot X_i^2}\right)^\cdot=0,
\ee
where dot means the total derivative with respect to $\tau$,
\be
s_i=\sqrt{l^2-x_i^2},
\ee
and
\be
\dot X_i=b(lv^0+x_iv^1)+(\dot x_i+\dot z-\dot rn)n.
\ee
Using Eqs. (140), we get
\be
\dot x_i=0,
\ee
\be
(-1)^ias_i-m_ix_i/s_i=0.
\ee
Eq. (144) yields that the quarks can not move along the straight-line
string.
Eq. (145) coincides with Eqs. (52) and (32).

It is not difficult to check that the Lagrangian (49), (46), (48) and
(51),
used in Sec. 3, gives exactly the same equations of motion (140).
(144) and
(145).



\begin{center}
{\large\bf Appendix B: The SQM Lagrangian in the configuration
space}\\
\end {center}

\vspace*{0.2cm}
Neglecting a total $\tau$-derivative, we can rewrite the SQM
Lagrangian (71)
in the form
\be
{\cal L}=-p\dot r-\pi\dot q-(i/2)\sum{\xi_i^M\dot\xi_{iM}}-H,
\ee
\be
H=c\left(J-K+i\sum{F_{ia}c_i^a\lambda_i}\right)+c_1pq+c_2\pi q,
\ee
where
\be
c_i^a=n^a\xi_i,\;\;\;c_i^5=\xi_i^5,
\ee
\be
n^0=p/m,\;\;\;n^1=\pi_1/\sqrt{-\pi_1^2},\;\;\;
n_{\mu}^2=\epsilon_{\mu\nu\rho\sigma}n^{0\nu}n^{3\rho}n^{1\sigma},
\ee
\be
n^3=q_p/\sqrt{-q_p^2},
\ee
\be
\pi_1^{\mu}=(g^{\mu\nu}-n^{0\mu}n^{0\nu}+n^{3\mu}n^{3\nu})\pi_{\nu},\
;\;\;
q_p^{\mu}=(g^{\mu\nu}-n^{0\mu}n^{0\nu})q_{\nu},
\ee
\be
c_i^{ab}=c_i^ac_i^b,\;\;\;c_{i,j}^{ab,ef}=c_i^{ab}c_j^{ef},
\ee
\be
J=\sqrt{-J^2}=L(1+t)+i\sum{c_i^{13}},\;\;\;
t=\frac{1}{2L^2}\sum{(c_{i,j}^{12,12}+c_{i,j}^{23,23})},
\ee
\be
L=\sqrt{q_p^2\pi_1^2},
\ee
\be
K=\bar lm-G(\bar l),\;\;\;m=G_{\bar l}(\bar l)
\ee
\be
F_{i0}=\bar l\sqrt{am_i/\bar l_i}\;,\;\;\;F_{i1}=(-1)^i\sqrt{am_i\bar
l_i}\;,\;
\;\;F_{i5}=-m_i.
\ee
Here $\bar l_i$ is given by Eq. (32) with substitution $\bar l$ for
$l$ where
$\bar l$ is a solution of the second Eq. (155).

The velocity variables are determined by the inverse Legendre
transformation
\be
\dot r=-H_{(p)},\;\;\;\dot q=-H_{(\pi)},
\ee
where index in brackets means the corresponding derivative. We can
use
constraints following from Eq. (147) after the differentiation. We
get
\be
\dot r=cl_0n^0+cy-c_1q,
\ee
\be
l_0=\bar l+\delta_0,\;\;\;\delta_0=-i\sum{F_{ia(m)}c_i^a\lambda_i},
\ee
\be
y=-J_{(p)}-i\sum{F_{ia}c_{i(p)}^a\lambda_i},\;\;\;yn^0=0,
\ee
\be
\dot q=c\alpha\pi_1+c\sqrt{-q^2}\;\gamma n^2-c_2q,
\ee
\be
\alpha=\sqrt{q_p^2\pi_1^2}\;(1-t).
\ee
\be
\gamma=i\sum{\left(\frac{1}{J}c_i^{13}+\frac{1}{L}F_{i1}c_i^2\lambda_
i\right)}.
\ee
Eqs. (146), (147), (158) and(161) yield
\be
{\cal L}=-c\left(G(\bar l)+\delta_0m+2Lt+
i\sum{(c_i^{13}+F_{ia}c_i^a\lambda_i)}\right)-(i/2)\sum{\xi_i^M\dot
\xi_{iM}}.
\ee

>From Eq. (161), we get
\be
c=b(1+\delta_b),
\ee
\be
\delta_b=t-(1/2)\gamma^2,
\ee
\be
v^1=(1-(1/2)\gamma^2)n^1+\gamma n^2.
\ee

Eqs. (158), (160) and (161) make it possible to find out $l$ and
$v^0$:
\be
l=\sqrt{\dot r_\perp}\;/b=l_0(1-\delta_l),
\ee
\be
\delta_l=-\delta_b+(1/2)\epsilon^2,
\ee
\be
\epsilon=(b_2-b_1\gamma)/l,
\ee
\be
b_1=\frac{1}{m}
\sum{\left[ic_i^{03}-\frac{1}{L}c_{i,j}^{02,12}+
i(F_{i0}c_i^1+F_{i1}c_i^0)\lambda_i\right]},
\ee
\be
b_2=\frac{1}{m}
\sum{\left[\frac{1}{L}(c_{i,j}^{01,12}-c_{i,j}^{03,23})+
iF_{i0}c_i^2\lambda_i\right]},
\ee
\be
v^0=(1+(1/2)\epsilon^2)n^0+\epsilon(-\gamma n^1+n^2).
\ee

Since $v^3=n^3$ on the constraints' surface, we can use Eqs. (167),
(173) and
(131) to get 
\be
v^2=(1+(1/2)\epsilon^2-(1/2)\gamma^2)n^2-\gamma n^1+\epsilon n^0.
\ee

The next step is to express all functions of $\bar l$ as functions of
$l$ using
Eqs. (159) and (169),
\be
\bar l=l+l_1,
\ee
\be
l_1=-\delta_0+l\delta_l,
\ee
and property of the Grassmann variables
\be
l_1^4=0.
\ee

Finally, we must express $c_i^a$ and their products through the
velocity
variables
\be
u_i^a=v^a\xi_i,\;\;\;a=0,1,2,3,\;\;\;u_i^5=\xi_i^5,
\ee
\be
u_i^{ab}=u_i^au_i^b,\;\;\;u_{i,j}^{ab,ef}=u_i^{ab}u_j^{ef},
\ee
where $v^a$ are given by Eqs. (50), (41), (131), (167), (173), (174)
and (175).
Using properties of the Grassmann variables, we obtain the Lagrangian
(45),
(46), (47) and (51)
\be
{\cal L}=-b\left(G(l)+i\sum{F_{ia}u_i^a\lambda_i}\right)-
(i/2)\sum{\xi_i^M\dot\xi_{iM}}+{\cal L}_{ss},
\ee
where
\be
{\cal
L}_{ss}=-b\left(A+i\sum{B_i\lambda_i}+C\lambda_1\lambda_2\right),
\ee
\be
A=i\sum{u_i^{13}}-K^{-1}u_{1,2}^{12,12},
\ee
\be
B_i=\left(-\frac{1}{lG'K}F_{i0}+\frac{l}{K^2}F'_{i0}\right)u_{i,j}^{0
12,12}-
\frac{i}{K}F_{i1}\left(u_{i,j}^{2,23}-\frac{2}{lG'K}u_{i,j}^{012,0123
}\right),
\ee
\ba
C&=&-\frac{1}{G''}\sum_{a,b}{F'_{ia}u_i^aF'_{jb}u_j^b}+
\left(\frac{1}{lG'}F_{i0}F_{j0}+\frac{1}{K}F_{i1}F_{j1}\right)u_{i,j}
^{2,2}+ 
\nonumber \\
&+&\frac{i}{K}S\biggl[\biggl(\frac{1}{G''}F'_{i1}-
\frac{l}{K}F_{i1}\biggr)F'_{j0}+\frac{1}{lG'}F_{i1}F_{j0}\biggr]
u_{i,j}^{2,023}+ \nonumber \\
&+ &\frac{i}{K}S\biggl[\biggl(\frac{1}{G''}F'_{i1}-
\frac{l}{K}F_{i1}\biggr)F'_{j1}-\frac{1}{K}F_{i1}F_{j1}\biggr]
u_{i,j}^{2,123}- \\
&- &2F_{i1}F_{j1}\biggl(\frac{1}{lG'K^2}u_{i,j}^{023,023,}+
\frac{1}{K^3}u_{i,j}^{123,123}\biggr)-\nonumber \\
&- &\biggr[\frac{2}{lG'K^2}F_{i1}F_{j1}+\frac{1}{lG'K}
\biggl(\frac{1}{G'}+\frac{l}{K}\biggr)SF_{i0}F'_{j0}+
\frac{l}{G''K^2}SF'_{i0}F''_{j0}+ \nonumber \\
&+ &\biggl(-\frac{2l^2}{K^3}+\frac{1}{G''K}\biggl(\frac{1}{K}
-\frac{lG'''}{KG''}-
\frac{2}{lG'}\biggr)\biggr)F'_{i0}F'_{j0}\biggr]u_{i,j}^{012,012}.
\nonumber 
\ea
Here all the functions depend on $l$, prime stands for derivative
with respect 
to $l$ and
\be
S X_{ij}=X_{ij}+X_{ji}.
\ee

The conserved spin constrains are
\be
\sum_{a}{F_{ia}u_i^a}+B_i-iC\lambda_j=0,
\ee
where $i=1$ or $2$ and $j\ne i$.

\vspace*{0.5cm}
\begin{center}
{\large\bf Appendix C: Comparison with experiment and potential 
quark model, parameters and predictions.}
\end{center}

\vspace*{0.2cm}
Comparison of SQM with experimental meson spectrum and a potential
quark model, SQM parameters and predictions are collected in Tables~
IV, V and VI below for different meson trajectories called in
correspondence with their lowest states. 

$q$ stands for quarks composing  mesons; 

$j^{PC}$ means $j^{P}$ for
mesons not having $C$- or $G$-parity;

$y$ (in GeV) is a main kinematical parameter of each meson, $y=al$,
where $l$ is a solution
of Eq.(61) and $a$ is the string tension. Knowing $y$, one can easily
calculate all the other parameters of the meson wave function. In the
classical approximation, $a/y$ is angular velocity of the string;

$m$ (in MeV) is SQM prediction for meson mass;

$m_{EXP}$ (in MeV) is experimantal meson mass from Ref.~[7] if no
reference
is indicated;

$\bullet$ indicates particles that appear in the Meson Summary
Tables~[7];

$m_P$ (in MeV) is a potential quark model prediction from Ref. [5];

$n$ is a meson name; and question marks stand for experimentally unknown
$j^{PC}$.

\setcounter{table}{3}
\pagebreak
\begin{table}
\caption{Vector trajectories (wave functions $\Psi_{jM,j-1,1},
P=C=(-1)^j, j_{\min}=1$).}
\begin{center}
\begin{tabular}{r|c|c|c|c|c|c|}
{} q     &  $j^{PC}$ & y & m & $m_{EXP}$ & $m_{P}$ & n
\\
\hline
$d\bar u$ & $1^{--}$ & 0.2450 & 771 & $\bullet 770.5\pm 0.8$ & 770 &
$\rho(770)$ \\
 & & & & $\bullet 781.94\pm 0.12$ & 780 & $\omega(782)$ \\
  \cline{2-7} 
 & $2^{++}$ & 0.4196 & 1319 & $\bullet 1318.1\pm 0.6$ & 1310 &
 $a_2(1320)$ \\
 & & & & $\bu 1275.0\pm 1.2$ &    & $f_2(1270)$ \\
  \cline{2-7}
 & $3^{--}$ & 0.5383 & 1692 & $\bu 1691\pm 5$ & 1680 & $\rho_3(1690)$
 \\
 & & & & $\bu 1667\pm 4$ &   & $\omega_3(1670)$ \\
\cline{2-7}
 & $4^{++}$ & 0.6346 & 1994 & $\bu 2060\pm 20$[9] & 2010 & 
$h/f_{4}$ (2050) \\
 & & & & $\bu 2010\pm 20$[10] & & $a_4(2040)$ \\
\cline{2-7}
  & $5^{--}$ & 0.7179 & 2256 & $2330\pm 35$[11] & 2300 &
  $\rho_5(2350)$ \\
\cline{2-7}
 & $6^{++}$ & 0.7924 & 2490 & $2510\pm 30$[12] & & $r/f_{6}$(2510) \\
\cline{2-7}
 & $7^{--}$ &  0.8603 & 2703 & & &  \\
\hline
$s\bar u$ & $1^-$ & 0.2600 & 893 & $\bu 891.66\pm 0.26$ & 900 & 
$K^*(892)^\pm$ \\
 & & & & $\bu 896.10\pm 0.28$ & & $K^*(892)^0$ \\
\cline{2-7}
 & $2^+$ & 0.4331 & 1418 & $ \bu 1425.6\pm 1.5$ & 1430 &
 $K^*_2(1430)^\pm$
 \\
 & & & & $ \bu 1432.4\pm 1.3$ & & $K^*_2(1430)^0$ \\
\cl
 & $3^-$ & 0.5509 & 1781 & $ \bu 1776\pm 7$ & 1790 & 
$K^*_3(1780)$ \\
\cl
 & $4^+$ & 0.6465 &  2077 & $ \bu 2045\pm 9$ & 2110 & $K^*_4(2045)$
 \\
\cl
 & $5^-$ & 0.7293 & 2334 &  $2382\pm 14\pm 19$ & & $K^*_5(2380)$ \\
\hline
$s\bar s$ & $1^{--}$ & 0.2758 & 1013 & $ \bu 1019.413\pm 0.008$ & 
1020 & $\phi(1020)$ \\
 & $2^{++}$ &  0.4469 & 1516 & $ \bu 1525\pm 5$ & 1530 & $f'_2(1525)$
 \\
 & $3^{--}$ & 0.5636 & 1870 & $\bu 1854\pm 7$ & 1900 & $\phi_3(1850)$
 \\
 & $4^{++}$ &  0.6586 & 2160 &   & 2200 & \\
 & $5^{--}$  & 0.7408 & 2413 &   & 2470 &  \\
 & $6^{++}$  & 0.8145 & 2640 &   &      &  \\
\hline
$c\bar u$ & $1^-$ & 0.3031 & 2008 & $\bu 2010.0\pm 0.5$ &
2040 & $D^*(2010)^\pm$ \\
 & & & & $\bu 2006.7\pm 0.5$ &   & $D^*(2007)^0$ \\
\cl
 & $2^+$ & 0.4996 & 2460 & $\bu 2458.9\pm 2.0$ & 2500 &
$D^*_2(2460)^0$ \\
 & & & & $\bu 2459\pm 4$ &  & $D^*_2(2460)^\pm$ \\
\cl
 & $3^-$ & 0.6264 & 2777 &  &  2830 &  \\
 & $4^+$ & 0.7269 & 3039 &   & 3110 &   \\
 & $5^-$ & 0.8127 & 3269 &  &     & \\
\hline
 $c\bar s$ & $1^-$ & 0.3244 & 2121 & $\bu 2112.4\pm 0.7$ & 2130 & 
$D^{*\pm}_s, ?^?$ \\
 & $2^+$ & 0.5163 & 2553 & $\bu 2573.5\pm 1.7$ & 2590 & 
$D_{sJ}(2573)^\pm, ?^?$ \\
  & $3^-$ & 0.6411 & 2861 &  & 2920 &  \\
 & $4^+$ & 0.7405 & 3118  &  &  3190 &  \\
 & $5^-$ & 0.8255 & 3344 &   &   &   \\
\hline
$c\bar c$ & $1^{--}$ & 0.3309 & 3097 & $\bu 3096,88\pm 0.04$ & 3100 & 
$J/\psi (1S)$ \\
 & $2^{++}$ & 0.6116 & 3557 & $\bu 3556.17\pm 0.13$ & 3550 &
 $\chi_{c2}(1P)$ \\
 & $3^{--}$ & 0.7412 & 3825 & & 3850 & \\
 & $4^{++}$ & 0.8415 & 4050 & & 4090 & \\
 & $5^{--}$ & 0.9267 & 4250 & &  & \\
\hline
$b\bar u$ & $1^-$ & 0.3629 & 5327 & $\bu 5324.9\pm 1.8$ & 5370 &
$B^*$ \\
 & $2^+$ & 0.5717 & 5716 & $5698\pm 12$ & 5800 &$B^*_J(5732), ?^?$ \\
 & $3^-$ & 0.7131 & 5994 & & 6110 & \\
 & $4^+$ & 0.8262 & 6224 & & 6360 & \\
 & $5^-$ & 0.9228 & 6426 & &  & \\
\hline
$b\bar s$ & $1^-$ & 0.3875 & 5432 & $5416.3\pm 3.3$ & 5450 &$B^*_s$
\\
 & $2^+$ & 0.5920 & 5803 & & 5880 & \\
 & $3^-$ & 0.7311 & 6073 & & 6180 & \\
 & $4^+$ & 0.8427 & 6298 & & 6430 & \\
 & $5^-$ & 0.9383 & 6497 & &  & \\
\hline
$b\bar c$ & $1^-$ & 0.5169 & 6489 &  & 6340 & \\
 & $2^+$ & 0.7292 & 6780 &  & 6770 & \\
 & $3^-$ & 0.8681 & 7003 &  & 7040 & \\
 & $4^+$ & 0.9781 & 7195 &  & 7270 & \\
 & $5^-$ & 1.0717 &7368  &  &  & \\
\hline
  & $1^{--}$ & 0.2274 & 9463 &$\bu 9460.37\pm 0.21$  & 9460
  &$\Upsilon
  (1S)$ \\
$b\bar b$  & $2^{++}$ & 0.8850 & 9912 & $\bu 9913.2\pm 0.6$  & 9900 &
$\chi_{b2}(1P)$ \\
 & $3^{--}$ & 1.0544 & 10106 &  & 10160 & \\
 & $4^{++}$ & 1.1791 & 10267 &  & 10360 & \\
 & $5^{--}$ & 1.2829 &10411  &  &  & \\
\hline
\multicolumn{7}{c}{$a=0.176\pm 0.002\;$GeV$^2$,}\\
\multicolumn{7}{c}{$m_s=224\pm 7,\;
m_c=1440\pm 10,\; m_b=4715\pm 20$,}\\
\multicolumn{7}{c}{$m_u=6.2\pm 0.2,\;
m_d=11.1\pm 0.4$,}\\
\multicolumn{7}{c}{$a_-(d\bar u)=a_-(s\bar u)=a_-(s\bar s)=0.88\pm
0.01$,}\\
\multicolumn{7}{c}{$a_-(c\bar u)=a_-(c\bar s)= A(c\bar c)=0.90,\;
B(c\bar c)=1.43$,}\\
\multicolumn{7}{c}{$a_-(b\bar u)=a_-(b\bar s)=a_-(b\bar c)=A(b\bar
b)=0.77,\;
B(b\bar b)=3.14$}\\
\end{tabular}
\end{center}
\end{table}

\begin{center}
\begin{table}
\caption{Pseudoscalar and pseudovector trajectories  (wave
functions $\Psi_{jMj0}$, $C=-P=(-1)^j$, $j_{\min}=0$ 
and $\Psi_{jMj1},$ $P=C=-(-1)^j$, $j_{\min}=1$,
or mixed states Eqs.~(100) for strange mesons).}
\begin{tabular}{r|c|c|c|c|c|c|}
{}q     &  $j^{PC}$ & y & m & $m_{EXP}$ & $m_{P}$ & n
\\
\hline
$d\bar u$ & $0^{-+}$ & 0.04311 & 138 & $ \bu 139.56995\pm 0.00035$ &
150 &
$\pi^\pm$\\
          &        &          &      &$\bu 134.9764\pm 0.0006$ &
          &
          $\pi^0$\\
\cl
 & $1^{+-}$ & 0.4006 & 1259 & $\bu 1229.5\pm 3.2$ & 1220 &
 $b_1(1235)$ \\
 &          &        &      & $\bu 1170\pm 20$ &      & $h_1(1170)$
 \\
\cline{2-2}\cline{5-7}
 & $1^{++}$ &    &    & $\bu 1230\pm 40$ & 1240 & $a_1(1260)$ \\
 &          &    &    & $\bu 1281.9\pm 0.6$    &  & $f_1(1285)$\\
\cline{2-7}
 & $2^{-+}$ & 0.5258 & 1653 & $\bu 1670\pm 20$ & 1680 & $\pi_2(1670)$
 \\
\cline{2-2}\cline{5-7}
 & $2^{--}$ &        &      &              & 1700 &   \\
\cl
 & $3^{+-}$ & 0.6247 & 1963 &        & 2030 &  \\
\cline{2-2}\cline{5-7}
 & $3^{++}$ &        &      &              & 2050 &   \\
\cl
 & $4^{-+}$ & 0.7093 & 2229 &     & 2330 &   \\
\cline{2-2}\cline{5-7}
 & $4^{--}$ &        &      &              & 2340 &   \\
\cl
 & $5^{+\mp}$ & 0.7847 & 2466 &     &  &   \\
\hline
$s\bar s$ & $0^{-+}$ & 0.09231 & 548 & $\bu 547.30\pm 0.12$ & 520 &
$\eta$ \\
\cl
 & $1^{+-}$ & 0.4307 & 1468 &     & 1470 &   \\
\cline{2-2}\cline{5-7}
  & $1^{++}$ &     &     & $\bu 1426.2\pm 1.2$ & 1480 & $f_1(1420)$
  \\
  &          &     &     & $1512\pm 4$ &    & $f_1(1510)$ \\
\cl
 & $2^{-+}$ & 0.5532 & 1838 &     & 1890 &   \\
\cline{2-2}\cline{5-7}
 & $2^{--}$ &        &      &              & 1910 &   \\
\cl
 & $3^{+-}$ & 0.6503 & 2135 &     & 2220 &   \\
\cline{2-2}\cline{5-7}
 & $3^{++}$ &        &      &              & 2230 &   \\
\cl
 & $4^{-+}$ & 0.7338 & 2391 &     & 2510 &   \\
\cline{2-2}\cline{5-7}
 & $4^{--}$ &        &      &              & 2520 &   \\
\cl
 & $5^{+\mp}$ & 0.8082       &   2621   &    &     &   \\
\hline
$c\bar c$ & $0^{-+}$ & 0.2219 & 2980 & $\bu 2979.8\pm 2.1$ & 2970 &
$\eta_c(1S)$ \\
\cl
 & $1^{+-}$ & 0.6075 & 3548 & $3526.14\pm 0.24$    & 3520 &
 $\eta_c(1P), ?^{??}$  \\
\cline{2-2}\cline{5-7}
  & $1^{++}$ &     &     & $\bu 3510.53\pm 0.12$ & 3510 &
  $\chi_{c1}(1P)$
  \\
\cl
  & $2^{-\pm}$ & 0.7385  & 3819  &    & 3840 & \\
\cl
 & $3^{+-}$ & 0.8395 & 4045 &     & 4090 &   \\
\cline{2-2}\cline{5-7}
  & $3^{++}$ &   &  &    & 4100 & \\
\cl
  & $4^{-\pm}$ & 0.9250    & 4246    &  &  & \\
\hline
  & $0^{-+}$ & 0.3361 & 9501 &    & 9400 & \\
\cl
$b\bar b$ & $1^{+-}$ & 0.8655 & 9892 &     & 9880 &   \\
\cline{2-2}\cline{5-5}\cline{7-7}
  & $1^{++}$ &     &     & $\bu 9891.9\pm 0.7$ &   & $\chi_{b1}(1P)$
  \\
\cl
 & $2^{-\pm}$ & 1.0357 & 10083 &     & 10150 &   \\
\cl
 & $3^{+\mp}$ & 1.1633 & 10245 &     & 10350 &   \\
\cl
 & $4^{-\pm}$ & 1.2690 & 10390 &     &  &   \\
\hline
$s\bar u$ & $0^{-}$ & 0.1209 & 494 & $\bu 493.677\pm 0.016$ & 470 &
$K^\pm$ \\
 &      &  &    & $\bu 497.672\pm 0.031$ &   & $K^0$ \\
\cl
 & $1^{+}$ & 0.4390 & 1436 & $\bu 1402\pm 7$ & 1380 & $K_1(1400)$ \\
\cline{3-7}
 & & 0.3978 & 1310 & $\bu 1273\pm 7$ & 1340 & $K_1(1270)$ \\
\cl
 & $2^{-}$ & 0.5576 & 1802 & $\bu 1816\pm 13$ & 1810 & $K_2(1820)$ \\
\cline{3-7}
 & & 0.5261 & 1704 & $\bu 1773\pm 8$ & 1780 & $K_2(1770)$ \\
\cl
 & $3^{+}$ & 0.6528 & 2097 &  & 2150 & \\
\cline{3-7}
 & & 0.6262 & 2014 & & 2120 & \\
\cl
 & $4^{-}$ & 0.7351 & 2352 &  & 2440 & \\
\cline{3-7}
 & & 0.7116 & 2279 & & 2410 & \\
\cl
 & $5^{+}$ & 0.8087 & 2582 &  &  & \\
\cline{3-7}
 & & 0.7875 & 2516 & &   & \\
\hline
$c\bar u$ & $0^{-}$ & 0.2366 & 1869 & $\bu 1869.3\pm 0.5$ & 1880 &
$D^\pm$ \\
 &      &  &    & $\bu 1864.6\pm 0.5$ &   & $D^0$ \\
\cl
& $1^{+}$ & 0.5228 & 2516 & $\bu  2422.2\pm 1.8$  & 2490
&$D_1(2420)^0$ \\
\cl
& $2^{-}$ & 0.6465 & 2828 &  &  & \\
\hline
$c\bar s$ & $0^{-}$ & 0.2491 & 1972 & $\bu 1968.5\pm 0.6$ & 1980 &
$D^\pm_s$ \\
\cl
& $1^{+}$ & 0.5350 & 2598 &$\bu 2535.35\pm 0.34\pm 0.5$   & 2570 &
$D_{s1}(2536)^\pm$ \\
\cl
& $2^{-}$ & 0.6576 & 2903 &  &  & \\
\hline
$b\bar u$ & $0^{-}$ & 0.3364 & 5279 & $\bu 5278.9 \pm 1.8$ & 5310 &
$B^\pm$ \\
       &         &        &      & $\bu 5279.2 \pm 1.8$ &   & $B^0$
       \\
\cl
& $1^{+}$ & 0.6153 & 5800 &  &  & \\
\cl
& $2^{-}$ & 0.7495 & 6067 &  &  & \\
\hline
$b\bar s$ & $0^{-}$ & 0.3501 & 5368 & $\bu 5369.3 \pm 2.0$ & 5390 &
$B_s^0$
\\
\cl
 & $1^{+}$ & 0.6290 & 5873 &$5853\pm 15$   &  &$B_{sJ}(5850), ?^?$ \\
\cl
 & $2^{-}$ & 0.7624 & 6135 &  &  & \\
\hline
$b\bar c$ & $0^{-}$ & 0.4411 & 6403 &$6400 \pm 390 \pm 130$ & 6270
&$B_c$ \\
\cl
 & $1^{+}$ & 0.7516 & 6814 &  &  & \\
\cl
 & $2^{-}$ & 0.8876 & 7036 &  &  & \\
\hline
\multicolumn{7}{c}{$a_0(d\bar u)=a_1(d\bar u)=-0.016,\;
a_0(s\bar s)=a_1(s\bar s)=-0.034$,} \\
\multicolumn{7}{c}{$a_0(c\bar c)=a_1(c\bar c)=-0.084,\;
a_0(b\bar b)=a_1(b\bar b)=-0.091$,} \\
\multicolumn{7}{c}{$a_0(s\bar u)=-0.10,\;\;
a_1(s\bar u)=0,\; d(s\bar u)=0.10$} \\
\multicolumn{7}{c}{$a_0(c\bar u)=a_1(c\bar u)=-0.30,\;
a_0(c\bar s)=a_1(c\bar s)=-0.27$,} \\
\multicolumn{7}{c}{$a_0(b\bar u)=a_1(b\bar u)=-0.55,\;
a_0(b\bar s)=a_1(b\bar s)=-0.51$,} \\
\multicolumn{7}{c}{$a_0(b\bar c)=-0.41\;$
(linear extrapolation between $a_0(b\bar u)$ and $a_0(b\bar b)$),}\\
\multicolumn{7}{c}{$d(c\bar u)=d(c\bar s)=
d(b\bar u)=d(b\bar s)=d(b\bar c)=0$} \\
\end{tabular}
\end{table}

\begin{table}
\caption{Scalar trajectories (wave functions 
$\Psi_{jM,j+1,1}$, $P=C=(-1)^j$, $j_{\min}=0$.)}
\begin{tabular}{r|c|c|c|c|c|c|}

{}q     &  $j^{PC}$ & y & m & $m_{EXP}$ & $m_{P}$ & n
\\
\hline
$d\bar u$ & $0^{++}$ & 0.3143 & 988 & $\bu 983.4\pm 0.9$ & 1090 &
$a_0(980)$\\
\cl
 & $1^{--}$ & 0.5073 & 1594 & $\bu 1700\pm 20$ & 1660 & $\rho(1700)$
 \\
 &                 &     &  &  $\bu 1649\pm 24$ &  &  $\omega(1600)$
 \\
\cl
 & $2^{++}$ &  0.6110 & 1920 & $1924\pm 14 [9, 13]$ & 2050 &
 $X(1920)$, $?^{??}$ \\
\cl
 & $3^{--}$  & 0.6979 & 2193 &    & 2370 & \\
\cl
 & $4^{++}$  & 0.7746 & 2434 &    &  & \\
\hline
$s\bar u$ & $0^{+}$ & (I) 0.4358 & (I) 1426 & $\bu 1429\pm 6$ & 1240
&
$K^*_0(1430)$\\
 &                   & (II) 0.3493 & (II)1162 &            &      &
 \\
\cl
& $1^{-}$ & (I) 0.5927 & (I) 1910 & $\bu 1717\pm 27$ & 1780 &
$K^*(1680)$\\
 &                   & (II) 0.5330 & (II)1726 &            &      &
 \\
\cl
& $2^{+}$ & (I) 0.6846 & (I) 2196 &    & 2150 & \\
&         & (II) 0.6339 & (II) 2038 &   &   & \\
\cl
& $3^{-}$ & (I) 0.7639 & (I) 2442 &    & 2460 & \\
 &   & (II) 0.7189 & (II) 2302 &    &  & \\
\cl
& $4^{+}$ & (I) 0.8352 & (I) 2664 &    &  & \\
&    & (II) 0.7943 & (II) 2537 &    &  & \\
\hline
$s\bar s$ & $0^{++}$ & 0.2726 & 1004 & $\bu 980\pm 10$ & 1360 &
$f_0(980)$\\
  & $1^{--}$ & 0.4922  & 1653  & $\bu 1680\pm 20$ & 1880 &
  $\varphi(1680)$\\
  & $2^{++}$ & 0.6018  & 1986  & $\bu 2011^{+60}_{-80}$ & 2440 &
  $f_2(2010)$\\
  & $3^{--}$ & 0.6918  & 2262  &   & 2540 & \\
  & $4^{++}$ & 0.7701  & 2505  &   &   & \\
\hline
$c\bar c$ & $0^{++}$ & 0.5357 & 3414 & $\bu 3417.3\pm 2.8$ & 3440 &
$\chi_{c0}(1P)$\\
  & $1^{--}$ & 0.7319  & 3805  & $\bu 3769.9\pm 2.5$ & 3820 &
  $\psi(3770)$\\
  & $2^{++}$ & 0.8360  & 4037  &   & 4090 & \\
  & $3^{--}$ & 0.9224  & 4240  &   &  & \\
\hline
 & $0^{++}$ & 0.8340 & 9860 & $\bu 9859.8\pm 1.3$ & 9850 &
 $\chi_{b0}(1P)$\\
$b\bar b$  & $1^{--}$ & 1.0663  & 10121  &   & 10140 & \\
  & $2^{++}$ & 1.1905  & 10282  & $\bu 10268.5\pm 0.4$  & 10350 &
  $\chi_{b2}(2P)$\\
  & $3^{--}$ & 1.2930  & 10426  &   &  & \\
\hline
\multicolumn{7}{c}{$a_+(d\bar u)=-0.88\;\;$,
(I)$a_+(s\bar u)=-1.59\;\;$, (II)$a_+(s\bar u)=-1.0$,} \\
\multicolumn{7}{c}{$a_+(s\bar s)=-0.52,\;\;
a_+(c\bar c)=-1.06,\; a_+(b\bar b)=-1.35$} \\
\end{tabular}
\end{table}
\end{center}



\begin{thebibliography}{99}
\bibitem{1}
L. D. Soloviev, Phys. Rev. D ${\bf 58}$, 035005 (1998).
\bibitem{2}
L. D. Soloviev, Teor. Mat. Fiz. ${\bf 116}$, 225 (1998).
\bibitem{3}
F. A. Berezin and M. S. Marinov, Ann. Phys. (N.Y.) ${\bf104}$, 336
(1977).
\bibitem{4}
L. Faddeev and R. Jackiw, Phys. Rev. Lett. ${\bf 60}$, 1692 (1988).
\bibitem{5}
S. Godfrey and N. Isgur, Phys. Rev. D ${\bf 32}$, 189 (1985).
\bibitem{6}
P. A. M. Dirac, {\it Lectures on Quantum Mechanics} (Yeshiva
University
New York, 1964).
\bibitem{7}
Particle Data Group, Caso {\it et al}., The European Physical
Journal,
C ${\bf 3}$, 1 (1998).
\bibitem{8}
S. Weinberg, Trans. NY Acad. Sci. ${\bf 38}$, 185 (1977).
\bibitem{9}
D. Alde {\it et al}., Phys. Lett. ${\bf 216B}$, 451 (1989); 
Sov. J. Nucl.Phys., ${\bf 49}$, 1021 (1989); Phys. Lett. ${\bf
241B}$, 600 
(1990).
\bibitem{10}
D. Alde {\it et al}, Yad. Fiz., ${\bf 59}$, 1027 (1996).
\bibitem{11}
D. Alde {\it et al}., Zeit. Phys., ${\bf 54}$, 553 (1992).
\bibitem{12}
F. Binon {\it et al}., Lett. Nuovo Cim., ${\bf 39}$, 41 (1984).
\bibitem{13}
G. M. Beladidze {\it et al}., Zeit. Phys., ${\bf 54}$, 367 (1992).
\end{thebibliography}
\end{document}